\documentclass[aps,pra, twocolumn, noshowpacs, floatfix]{revtex4-2}

\usepackage{wrapfig}
\usepackage{amsfonts}
\usepackage{amssymb}
\usepackage{graphicx}
\usepackage{amsmath}
\usepackage{amsmath}
\usepackage[english]{babel}
\usepackage[mathscr]{euscript}
\usepackage{color}
\usepackage{epstopdf}
\usepackage{footnote}
\usepackage{isotope}
\usepackage{ulem}
\usepackage{float}

\usepackage[export]{adjustbox}

\begin{document}

\title{Hamiltonian non-Hermicity: accurate dynamics with the multiple Davydov D$_2$ Ans\"atze}

\author{Lixing Zhang$^{1,2}$, Kaijun Shen$^{1}$, Yiying Yan$^{3}$, Kewei Sun$^{4}$, Maxim F. Gelin$^{4}$, and Yang Zhao$^{1}$\footnote{Electronic address:~\url{YZhao@ntu.edu.sg}}}

\affiliation{$^{1}$\mbox{School of Materials Science and Engineering, Nanyang Technological University, Singapore 639798, Singapore}\\
$^{2}$\mbox{Department of Chemistry and Biochemistry, University of California Los Angeles, Los Angeles, CA 90095, USA}\\
$^{3}$\mbox{School of Science, Zhejiang University of Science and Technology, Hangzhou 310023, China}\\
$^{4}$\mbox{School of Science, Hangzhou Dianzi University, Hangzhou 310018, China}\\
}

\begin{abstract}
We examine the applicability of the numerically accurate method of time dependent variation with multiple Davydov Ans\"atze (mDA) to non-Hermitian systems.  {As illustrative examples, three systems of interest have been studied, a non-Hermitian system of dissipative Landau-Zener transitions, a non-Hermitian multimode Jaynes-Cummings model, and a dissipative Holstein-Tavis-Cummings model, all of which are shown to be effectively described by the mDA method}. Our findings highlight the versatility of the mDA as a powerful numerical tool for investigating complex many-body non-Hermitian systems, which can be extended to explore diverse phenomena such as skin effects, excited-state dynamics, and spectral topology in the non-Hermitian field.
\end{abstract}

\date{\today}
\maketitle

\section{introduction}

The unconventional realm of non-Hermitian physics~\cite{Ash, Kaw, Ber, CQW}, stemming from a novel, mainly phenomenological framework of addressing quantum open-system dynamics, offers an alternative avenue for investigating the behaviors of open quantum systems and the fascinating topic of quantum dissipation. 
Non-Hermitian physics has spawned many novel phenomena, such as complicated bandgap definition~\cite{Shenprl, Kawprl, Huprl}, topological phase transition~\cite{Zhang, Long1, Aoprl, Long2, Long3, Zhaophy}, bulk-boundary correspondence~\cite{Kunstprl, Zirn}, and non-Hermitian skin effect~\cite{Sark, Clae, Kunstprb, Ochi, ccls1}. On the other hand, the description of various decoherence and lifetime processes in terms of non-Hermitian Hamiltonians has long been a method of choice in photophysics and spectroscopy\cite{Faisal, Jindra} and, furthermore, the effective non-Hermitian Hamiltonians can be constructed for driven   {quantum systems} \cite{Dalibard}.

Traditional approaches to quantum physics often assume isolation, where fundamental principles are based on the premise that the object of interest operates in a closed system without interactions with the environment. The energy operator of a quantum closed system is typically described by a Hermitian operator known as a Hamiltonian, represented by a Hermitian matrix. This depiction simplifies the physical model and highlights the conservation of energy within the system. However, in reality, many physical systems are open and susceptible to environmental influences.

While various methods exist to incorporate bath effects in the system dynamics, the non-Hermitian approach is unique in the semi-empirical manner in which it disrupts the well-known Hamiltonian Hermiticity in integrating environmental influences~\cite{natphys_2018,ref2}. Consequently, a novel formalism based on non-Hermitian Hamiltonians brings about energy and particle number non-conservation. Dissipation, stemming from environmental effects, stands out as a prevalent and inevitable consequence.

{Despite dissipation can be engineered in particular systems~\cite{Frank2020}, it is still viewed as a negative effect in various fields}, as it often diminishes the functionality of devices used for sensing, computation, and communication. However, non-Hermitian physics challenges this perspective by revealing unconventional physical effects associated with dissipation or a strategic interplay of dissipation and amplification. Particularly in open systems with gain or loss, the unique properties of non-conserved energy and non-unitary evolution result in complex eigenvalues and introduce exotic features to the eigenstates of the systems. 

One important application of non-Hermitian Hamiltonians is represented by cavity quantum electrodynamics (QED)~\cite{CR}, where cavity photon dissipation is modeled by adopting a complex photon frequency~\cite{McT}. In cavity materials science, the interaction between light and matter inside optical cavities leads to the formation of polaritonic states, which are hybrid states combining photonic and excitonic (matter) components. In a realistic experimental setup, cavity photons have a finite lifetime inside the cavity, leading to dissipative effects that are inherently non-Hermitian. 
To accurately describe these systems, one can phenomenologically incorporate various sources of dissipation into a non-Hermitian Hamiltonian, responsible for spectroscopic broadening of the light-matter eigenspectrum. Strong coupling in cavity QED is commonly defined when the matter-cavity coupling strength $g_c$ is much larger than the cavity loss rate $\kappa$ (assuming the matter de-excitation rate is much smaller than $\kappa$). Under these conditions, cavity loss described by the non-Hermitian Hamiltonian plays a significant role in the system's dynamics and influences polariton photochemistry significantly~\cite{Fre1, Fre2, Ant, Mar}.

Recent investigations into non-Hermitian Hamiltonians have revealed significant effects of cavity loss on polaritonic dynamics. One notable impact is the reduction of excited-state populations with photonic character, which alters the behavior of polaritonic states \cite{Koe}. Additionally, cavity loss can enhance the rates of photochemical reactions \cite{Tor} and protect molecules from photodamage by decreasing the time photoexcited molecules spend in nuclear configurations prone to damage~\cite{Fel}. However, when multiple excitation manifolds are accessible, the influence of cavity loss becomes more complex~\cite{Dav}. This complexity underscores the need for detailed, accurate calculations of non-Hermitian dynamics to optimize cavity parameters for controlling photochemical reactions.

Time dependent variation with multiple Davydov Ansatz (mDA) is an accurate method to unveil the dynamics of a complex many-body quantum system by solving the time-dependent Schr\"odinger equation (TDSE) numerically ``exactly"~\cite{Zhao1, JCP_P}. By expressing the bosonic part of the wave function with an overcomplete basis of coherent-state superpositions, mDA can capture the intricate dynamics for all multi-species bosonic degrees of freedom (DOF). Moreover, in the mDA algorithm, computational complexity only scales linearly with respect to the number of bosonic DOFs, which further empowers mDA to be highly adaptive to a diverse array of systems. 
For examples, mDA has been successfully applied to a variety of systems in chemical physics, condensed matter physics, cavity QED, nonlinear spectroscopy and many-body quantum dynamics~\cite{D2_1, Sun1, Shen1, Zheng1, D2_2, D2_3}. 

Despite the success of mDA in Hermitian systems, till date, it has not yet been adopted in the novel regime of non-Hermitian systems. In the remainder of this paper, we showcase the application of mDA to three particular non-Hermitian systems of interest. The first is a dissipative Landau-Zener (LZ) system with a tunable non-Hermitian term. Proven by the Kibble-Zurek theory~\cite{KZ_LZ}, the non-Hermitian effects in the LZ system is related to various of parameters including relaxation time, temperature and quenching time. The second is a non-Hermitian, multimode Jaynes-Cummings (JC) model, which usually describes  a two-level system interacting with lossy cavities or waveguides~\cite{Lzy2015,Lzy2016,ZJY2024}. In the third case, we use a non-Hermitian term to describe realistic experimental environment in a dissipative Holstein-Tavis-Cummings (HTC) model, which is a model that describes, e.g., cavity singlet fission processes with Kerr effect. In addition, we benchmark the results generated by mDA against  numerically exact solutions.

\section{METHODOLOGY}

\subsection{The multi-D$_2$ Ansatz and the Time Dependent Variational Principle}

The governing equation for the time evolution of quantum systems, the TDSE can be written as follow:

\begin{eqnarray}\label{EQ9}
i\frac{\partial}{\partial t}|\Psi(t) \rangle = \hat{H} |\Psi(t) \rangle
\end{eqnarray}

For non-Hermitian systems, {the bra ($\langle \langle n|$)  and ket ($| n\rangle$) eigenstates of the Hamiltonian} are not Hermitian conjugates. This leads to differences in the time evolution of $\langle \langle \Psi(t)|$ and $|\Psi(t) \rangle$ (i.e., $\langle \langle \Psi(t)| \neq |\Psi(t) \rangle^\dagger $). In this paper, we only  focus on the time evolution $|\Psi(t) \rangle$, {which is sufficient for the evaluation of all quantum observables.}

For complex systems, the complete solution of TDSE is almost impossible to obtain. Therefore, we employ the multiple Davydov D$_2$  {(mD2)} Ansatz, also known as the multi-D$_2$ Ansatz, to variationally solve the TDSE. For the three systems considered in this work, the mD2 Ansatz can be written as:
{\begin{eqnarray}\label{EQ9}
{|{\rm D}_{2}^M(t)}\rangle& =& \sum_{s}^{N_s}| s \rangle \sum_{n=1}^{M} A_{ns}(t)e^{(\sum_k^{N_b} \alpha_{nk}(t)b_k^{\dagger} - {\rm H. c.})}|\{0_{k}\}\rangle   \nonumber \\
\end{eqnarray}}
Here, $N_s$ is the total number of spin DOF, and $| s \rangle$ represents a spin state with spin $s$. Similarly, $N_b$ is the total number of bosonic DOF, and $|\{0_{k}\}\rangle$ is the multi-mode vacuum state. The multiplicity $M$ is the number of single D$_2$ Ans\"atze in the mD2 Ansatz. Each single D$_2$ Ansatz is characterized by a group of independent variational parameters: amplitude $A_{ns}(t)$ and displacement $\alpha_{nk}(t)$. This expresses the boson part of the wavefunctions through a linear combination of coherent states that have different displacements. They form an overcomplete basis that is capable of {reproducing} accurate {coupled spin-boson dynamics} at sufficiently large $M$.

Following the Lagrange's principle, the Ansatz can be used to solve the TDSE variationally. We construct the Lagrangian in the following form ($\hbar$ = 1)~\cite{Kra1, Bro, Yuan1}:
\begin{eqnarray}\label{EQ11}
L&=& \langle {\rm D}_{2}^{M}(t)|(\frac{d}{dt}+i \hat{H})|{\rm D}_{2}^{M}(t)\rangle
\end{eqnarray}

To obtain the time trajectory of the variational parameters, the action functional $S = \int_{t_1}^{t_2} L\ dt$ {has to} be minimized. This can be enforced by the Euler-Lagrange (EL) equation:
\begin{align}\label{EQ10}
\frac{d}{dt}\frac{\partial L}{\partial{\dot{u}}_n^\ast}-\frac{\partial L}{\partial u_n^\ast}=0,~ {{u}}_n \in [A_{ns}, \alpha_{nk}]
\end{align}
This yields the equations of motion (EOM), which are a set of coupled differential equations. The EOMs can be solved via the 4$^{th}$ order Runge-Kutta method.

\subsection{Observables}

For an arbitrary operator $Q$, its expectation value in the mD2 basis can be written as:
\begin{eqnarray}\label{EQ14}
&&\langle Q\rangle (t) = \langle{{\rm D}_{2}^{M}(t)}|Q|{{\rm D}_{2}^{M}(t)}\rangle
\end{eqnarray}

When $\hat H$ is non-Hermitian, the time evolution becomes non-unitary. This is characterized by a non-unitary normalization factor, which can be written as:

\begin{eqnarray}\label{EQ12}
N_f(t) = \langle D_2^M(t) | D_2^M(t) \rangle = \sum_{m,n}^{M}\sum_s^{N_s} A_{ms}^{\ast}(t){A}_{ns}(t)S_{mn} \nonumber \\
\end{eqnarray}
with:
\begin{eqnarray}\label{EQ13}
S_{mn} &=& {\rm exp}\left[\sum_k^{N_b}\alpha_{mk}^{\ast}\alpha_{nk}-\frac{1}{2}(\lvert{\alpha_{mk}}\lvert^2+\lvert{\alpha_{nk}}\lvert^2)\right]\nonumber \\
\end{eqnarray}

\section{RESULTS AND DISCUSSION}\label{RD}

\begin{figure}[t]
\centerline{\includegraphics[width=90mm]{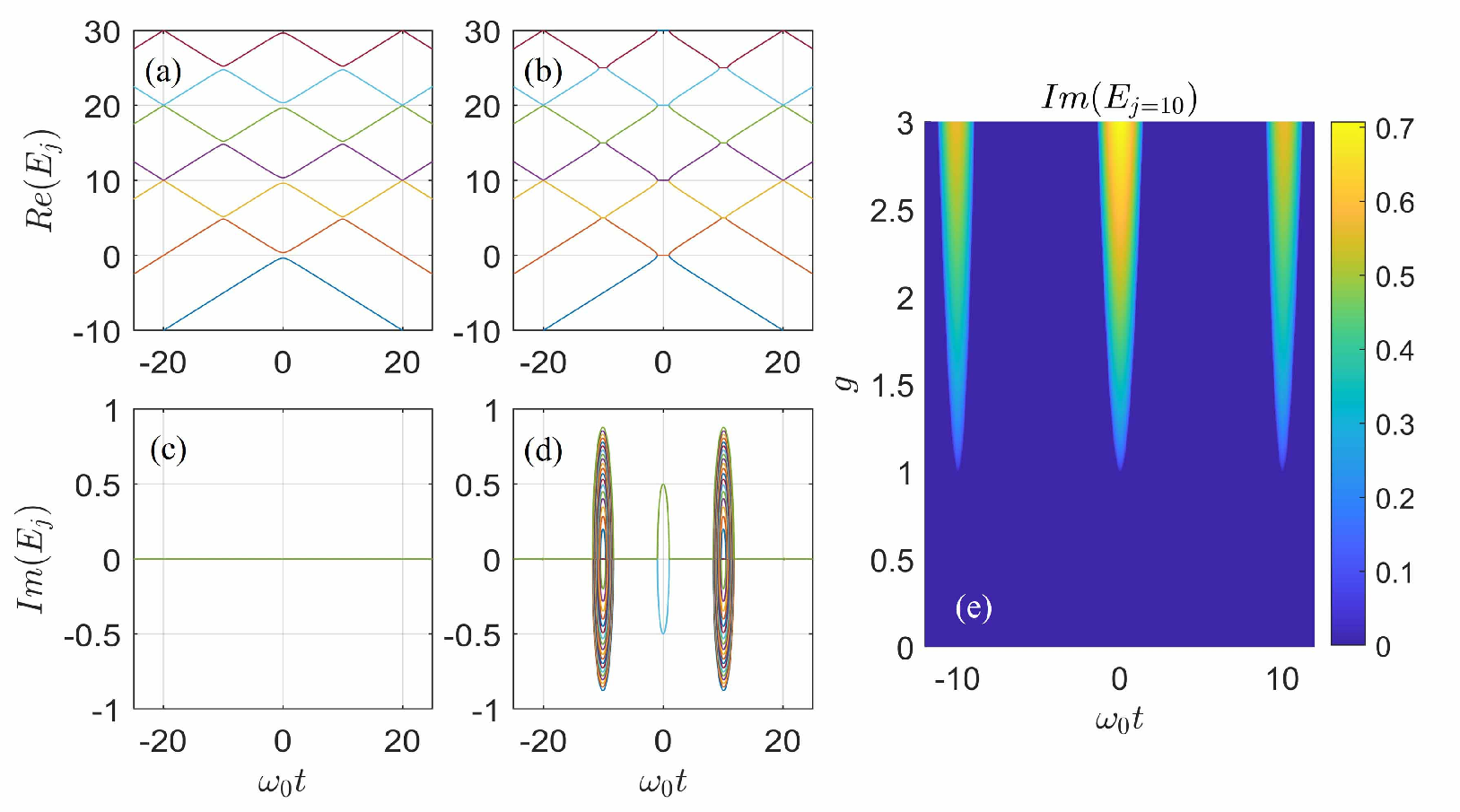}}
\caption{The eigenvalues of ${\hat H}$ with $N = 1$ . (a) and (c): The real and imaginary part of the eigenvalues when $g$=0.5. (b) and (d): The real and imaginary part of the eigenvalues when $g$=2. (e): Contour plot illustrating the maximum imaginary part of the eigenvalues over time and $g$. A distinct demarcation is noticeable at $g = 1$. Remaining parameters are as follows:  {$v = 0.5\omega_0^2$}, $\lambda$ = 0.2$\omega_0$, $\omega$ = 10$\omega_0$ and $\Delta$ = 0.5$\omega_0$ ($\omega_0$ is the unit frequency). }
\label{fig1}
\end{figure}

\subsection{Dynamics of the non-Hermitian LZ transitions}\label{Om}

A non-Hermitian LZ transition may be described by a non-Hermitian Hamiltonian as follows:

\begin{eqnarray}\label{EQ1}
{\hat H}_{\rm NLZ} &=&
 \begin{bmatrix}
     vt/2 & ~\Delta\\
     \Delta(1-g) &~ -vt/2 \\
 \end{bmatrix}
\end{eqnarray}
where $v$ is the scanning velocity, $\Delta$ is the tunneling strength between two levels, and $g$ is the strength of non-Hermitian level coupling. In particular, when $g = 0$, the Hermitian LZ problem is recovered.

In addition to the non-Hermitian coupling of electronic DOFs, we also consider the non-Hermitian interaction between spin and an Ohmic heat bath. The Hamiltonian for the total system can be written as:
\begin{eqnarray}\label{EQ1}
{\hat H} &=& {\hat H}_{\rm NLZ} + \sum_k^{N_b} \omega_k{\hat b_k^\dagger}{\hat b_k} +
 \begin{bmatrix}
	0   &~ 1 \\
   1-g &~ 0 \\
 \end{bmatrix}
\sum_k^{N_b} \frac{\lambda_k}{2}({\hat b_k^\dagger}+{\hat b_k}) \nonumber \\
\end{eqnarray}
Here, $\hat{b_k^\dagger}$($\hat{b_k}$) is the creation(annihilation) operator of the $k^{th}$ mode of the quantum bath. With a total number of $N_b$ bath modes, $\omega_k$ and $\lambda_k$ are the frequencies and the coupling strengths of the $k^{th}$ mode, respectively. The bath modes are characterized by an Ohmic-type spectral density function:
\begin{equation}\label{EQ6}
J(\omega)=\sum_{k}^{N_b}(\lambda_k)^{2}\delta(\omega -\omega _{k})=2\alpha \omega e^{-\omega/\omega_c}
\end{equation}
 where $\alpha$ is the strength of spin-bath coupling, and $\omega_c$ is the cut-off frequency.

\begin{figure}[t]
\centerline{\includegraphics[width=95mm]{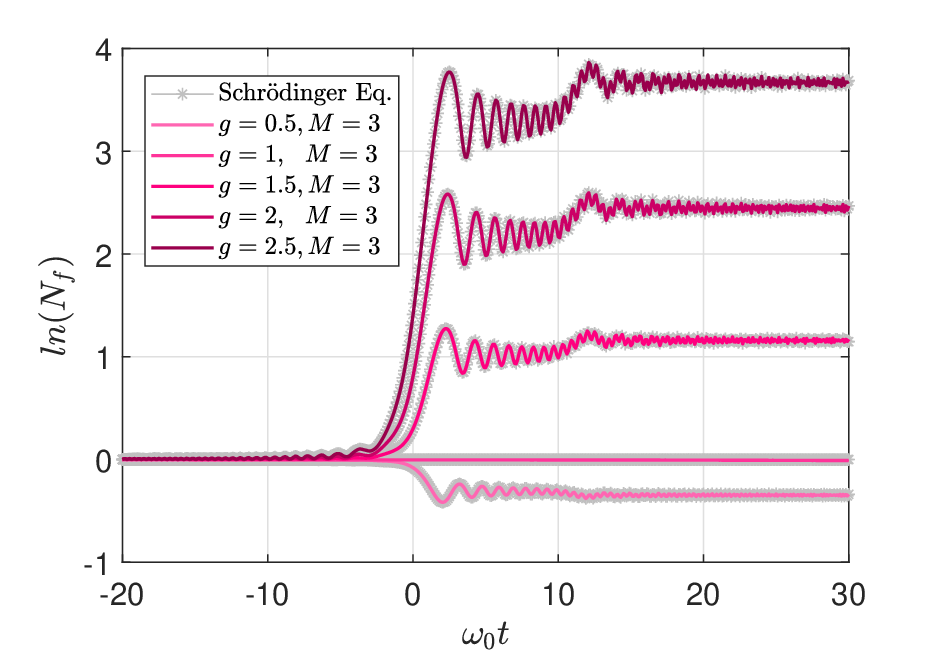}}
\caption{Time evolution of the norm of the wavefunction caculated by the TDSE (dotted grey line) and mD2 (red line, deeper colors represent higher magnitude of $g$) when $g$ is changed from $0.5$ to $2.5$. The rest of the parameters are set as: $\Delta = 0.5\omega_0$, $\lambda = 0.2\omega_0$, $\omega = 10\omega_0$, and $v = 0.5\omega_0^2$.}
\label{fig2}
\end{figure}

To get a physical picture of the problem, we first investigate the energy spectrum of $\hat H$ when $N_b =1$. In Fig.~\ref{fig1}, we present the effect of the non-Hermitian term $g$ on the energy spectrum of the problem. The magnitude of $g$ separates the problem into two phases marked by the emergence of imaginary eigenvalues. When $g = 0.5$, notwithstanding $\hat H$ becomes non-Hermitian, its eigenvalues, as depicted in Fig.~\ref{fig1}(a) and (c), are real across all time points, while the renowned avoided crossings remain observable. However, when $g = 2$, as shown in Fig.~\ref{fig1}(b) and (d), while the energy spectrum remains predominantly real across most time points, complex eigenvalues emerge as energy levels intersect, supplanting the previously observed avoided crossings. In Fig.~\ref{fig1}(e), by plotting the maximum imaginary eigenvalue with respect to time and $g$, it becomes evident that the transition point between these two cases is marked by $g=1$, which is known as the Exceptional Point (EP). When $g<1$, the eigenvalues of $\hat H$ remain real at any time points. Conversely, for $g>1$, complex eigenvalues emerge, {marking the crossing of EP}.

\begin{figure} 
\centerline{\includegraphics[width=100mm]{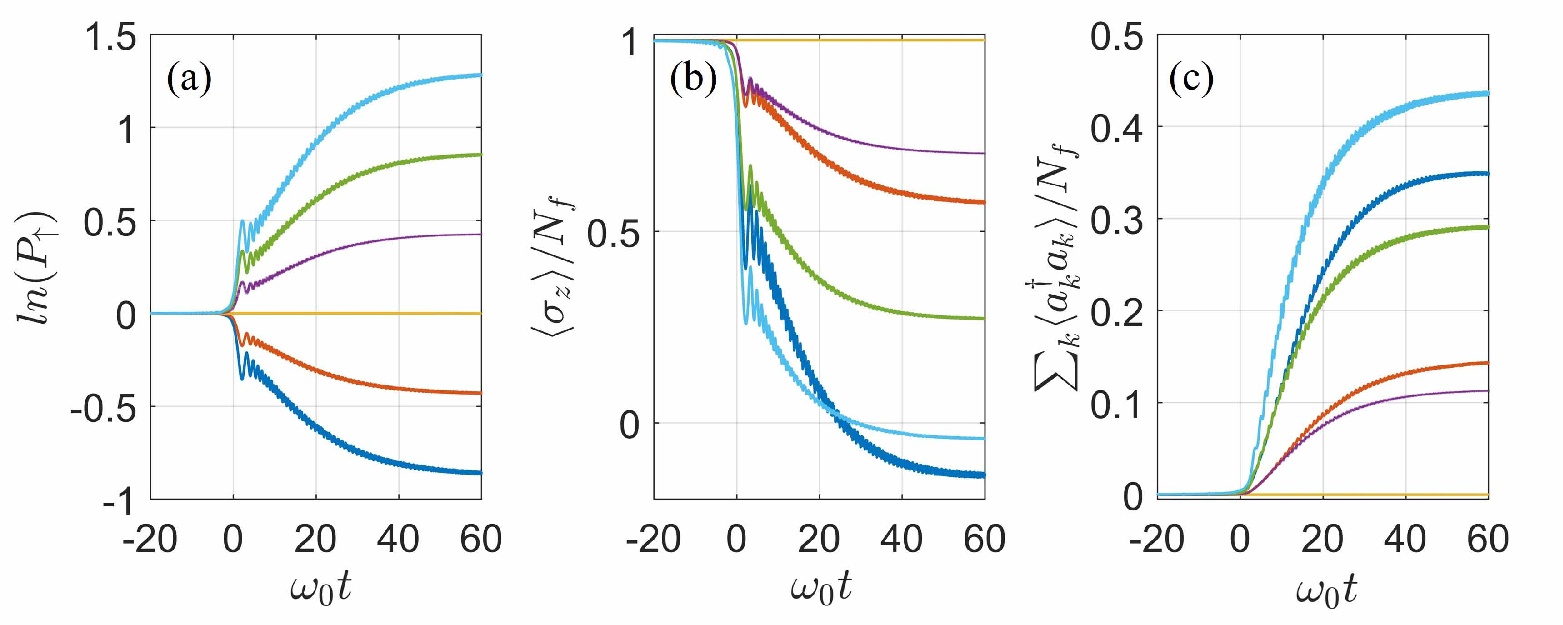}}
\caption{Time evolution of different observables calculated by mD2 with $M$=5. Different colors of lines indicate different magnitudes of $g$: navy blue: $g$=0; red: $g$=0.5; yellow: $g$=1; purple: $g$=1.5; green: $g$=2; sky blue: $g$=2.5   (a). Time evolution of the unnormalized upper- state population in logarithmic scale. (b). Time evolution of the normalized spin polarization. (c). Time evolution of the total bath phonon number. The rest of the parameters are set as: $\Delta = 0.5\omega_0$, $\lambda = 0.2\omega_0$, $\omega = 10\omega_0$, and $v = 0.5\omega_0^2$. }
\label{fig3}
\end{figure}

{We note that the mD2 Ansatz remains numerically accurate regardless of whether the eigenvalue of $\hat H$ is real or complex}. To compare the results produced by the mD2 Ansatz and by solving the TDSE, in Fig.~\ref{fig2}, we consider $N=1$ and plot the normalization factor $N_f$ against time in logarithmic scale across a range of values for $g$ spanning from $0.5$ to $2.5$. The colored lines (mD2 Ansatz with $M$=3) completely overlap with the grey dotted lines (TDSE), {hence the validity of the mD2 Ansatz is confirmed for ${\hat H}_{\rm NLZ}$}. Notably, at EP where $g=1$, $N_f$ remains unitary throughout the evolution. Unlike the unitary evolutions in Hermitian physics, this phenomenon arises from the absence of tunneling strength between $|\uparrow\rangle$ and $|\downarrow\rangle$. With the system initialized on $|\uparrow\rangle$, the wave function remains invariant throughout the evolution.

To further examine the applicability of the mD2 Ansatz for more complicated systems, in Fig.~\ref{fig3},  we increase $N_b$ to $60$, and showcase the time evolution of various observables across different strengths of non-Hermitian level coupling calculated by the mD2 Ansatz. While the parameters $\Delta = 0.2\omega_0$, $\alpha = 0.002$, and $\omega_c = 10\omega_0$ are held constant, the parameter $g$ is varied from $0$ to $2.5$ across three different regimes: {the Hermitian regime, real-eigenvalue regime, and imaginary-eigenvalue regime}. In Fig.~\ref{fig3}a, the time evolution of the unnormalized population is presented in logarithmic scale. Clearly, $P_{\uparrow}(t)$ increases with $g$ monotonically. The origin of such phenomenon arises from the non-Hermitian nature of the Hamiltonian. In Fig.~\ref{fig3}b and Fig.~\ref{fig3}c, we plot the time evolution of normalized observables, which are obtained by dividing the expectation value of unnormalized observables by the norm of the wavefunction. {When $g \leq 1$}, an increase in $g$ leads to a decrease in both the normalized spin polarization ($\langle \sigma_z \rangle/N_f$) and the total phonon number of the bath ($\sum_k\langle a_k^\dagger a_k \rangle/N_f$). This suggests that the effect of non-Hermicity in this regime can be roughly interpreted as a reduction in the effective spin-bath coupling strength. Similarly, {when $g \geq 1$}, an increase in $g$ results in an increase in both the normalized spin polarization and the total phonon number of the bath, implying an increase in the effective spin-bath coupling strength. Furthermore, a noticeable change in the gradient of the lines can be observed in  Fig.~\ref{fig3}b. This indicates that strength of the non-Hermitian level coupling also changes the effective spin-bath coupling.

\subsection{Dynamics of the non-Hermitian multi-mode JC model}

\begin{figure*}
\includegraphics[width=1\columnwidth]{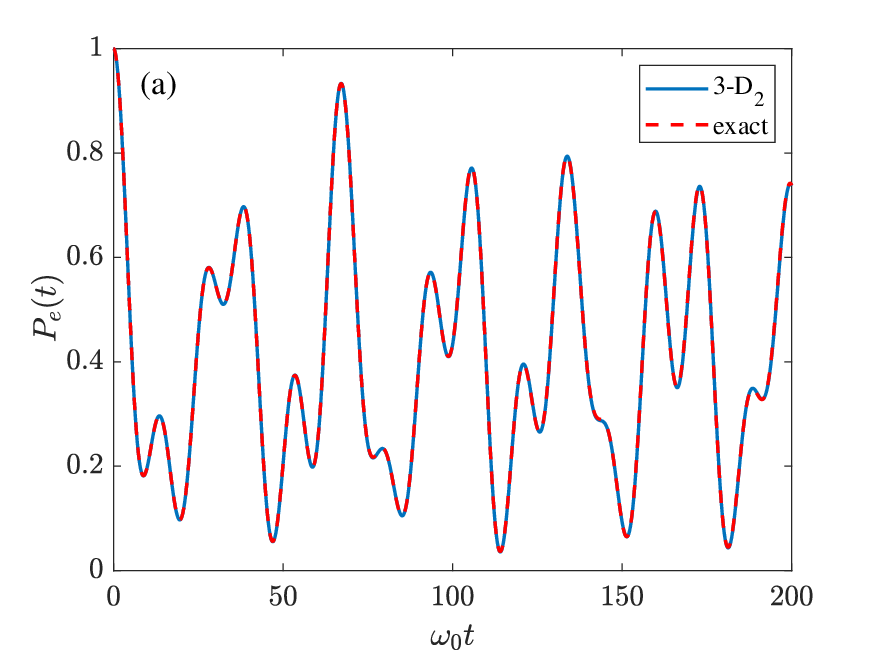}
\includegraphics[width=1\columnwidth]{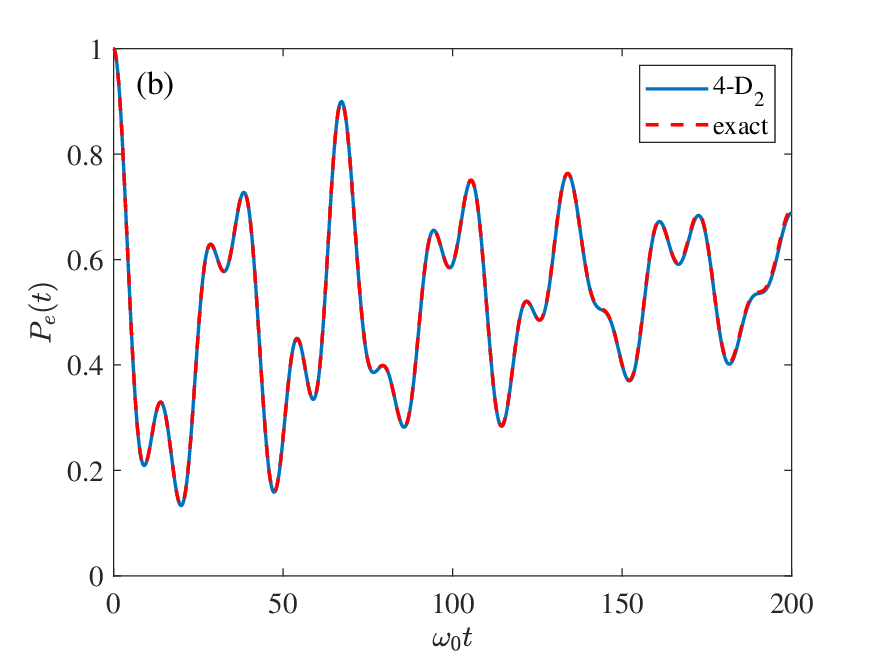}

\caption{Excited-state population versus time calculated by the mD2 (blue line) and TDSE (dashed red line) for $N_b=3$, $\omega_{k}/\omega_{0}=(1.0,1.2,1.3)$,
$g_{k}=0.2\omega_{0}$, $\gamma=10^{-2}\omega_{0}$, and (a) $\kappa_{k}=\gamma$,
and (b) $\kappa_{k}=5\gamma$. Ansatz multiplicity $M=3$ and 4 in the left and right panel, respectively.}\label{fig1yyy}

\end{figure*}

Our next model to consider is a non-Hermitian version of the much-studied, paradigmatic JC model.
The Hamiltonian of a non-Hermitian, multimode JC model reads
\begin{eqnarray}
H&=&\frac{1}{2}\omega_{0}\hat{\sigma}{}_{z}-\frac{i\gamma}{2}\hat{\sigma}_{+}\hat{\sigma}_{-}+\sum_{k}\left(\omega_{k}-\frac{i\kappa_{k}}{2}\right)\hat{a}_{k}^{\dagger}\hat{a}_{k}\nonumber\\
& &+\sum_{k}\frac{g_{k}}{2}(\hat{\sigma}_{-}\hat{a}_{k}^{\dagger}+\hat{\sigma}_{+}\hat{a}_{k})
\end{eqnarray}
where $\hat{a}_k^{\dagger}$ ($\hat{a}_k$) represents the creation (annihilation) operator for the $k^{\rm th}$ mode of cavity, and $\hat{\sigma}^{+}=|e\rangle\langle g|$
($\hat{\sigma}^{-}=|g\rangle\langle e|$) are the raising (lowering) Pauli operators.
$\omega_0$ is the transition frequency of the two-level system. $\omega_k$ are the frequencies of cavity modes. $\gamma$ and $\kappa_{k}$ denote the decay rates of the excited state of the two-level
system and the cavity modes, respectively. Time evolution of this model can be exactly solved
due to its symmetry, which conserves the total excitation number $\hat{N_e}=\hat{\sigma}_{+}\hat{\sigma}_{-}+\sum_{k}\hat{a}_{k}^{\dagger}\hat{a}_{k}$ {in the absence of non-Hermicity}, and thus provides an ideal testbed to check the accuracy of the present variational approach.
We consider a single-excitation subspace in which the time evolution can be described by
the following Aansatz
\begin{equation}
|\Psi(t)\rangle=c_{e}(t)|e,\{0_{k}\}\rangle+\sum_{k}c_{gk}(t)|g,\{1_{k}\}\rangle.
\end{equation}
Here $c_{e}(t)$ and $c_{gk}(t)$ are the probability amplitudes. $|e,\{0_{k}\}\rangle=|e\rangle\otimes|\{0_{k}\}\rangle$
and $|g,\{1_{k}\}\rangle=|g\rangle\otimes|\{1_{k}\}\rangle$ where
$|e\rangle$ and $|g\rangle$ are the excited and ground states of
the two-level system, respectively. $|\{0_{k}\}\rangle$ is the multi-mode
vacuum state and $|\{1_{k}\}\rangle$ implies that one photon occupies
the mode $k$ and the remaining modes are empty. From the TDSE, we obtain EOMs for the coefficients: 
\begin{equation}
i\frac{d}{dt}c_{e}(t)=\frac{\omega_{0}-i\gamma}{2}c_{e}(t)+\sum_{k}\frac{g_{k}}{2}c_{gk}(t),
\end{equation}
\begin{equation}
i\frac{d}{dt}c_{gk}(t)=\left(\omega_{k}-\frac{\omega_{0}}{2}-i\frac{\kappa_k}{2}\right)c_{gk}(t)+\frac{g_{k}}{2}c_{e}(t).
\end{equation}
Provided $|\Psi(0)\rangle=|e,\{0_{k}\}\rangle$, the EOMs can be numerically solved and the population of the excited state of the two-level system can be computed as
\begin{equation}
P_{e}(t)=\frac{|c_{e}(t)|^{2}}{|c_{e}(t)|^{2}+\sum_{k}|c_{gk}(t)|^{2}}\label{eq:pet}
\end{equation}
This is referred to as a numerically exact solution and is used to benchmark the variational results. {It is noted that a non-Hermitian Hamiltonian can also be constructed from the Lindblad master equation for a dissipative two-level system coupled to a lossy multimode cavity by neglecting quantum jump terms. The present time-dependent variational approach may be further combined with the quantum trajectory approaches to simulate the open quantum systems described by the Lindblad master equation\cite{Liam2024}.}

Fig.~\ref{fig1yyy} showcases the comparison between the numerically exact and
variational solutions for $N_b=3$, $g_{k}=0.2\omega_{0}$, $\omega_{k}/\omega_{0}=(1.0,1.2,1.3)$,
$\gamma=10^{-2}\omega_{0}$, and two sets of $\kappa_{k}$.  The mD2 Ansatz multiplicity, $M$, is 3 and 4 in the left and right panel of Fig.~\ref{fig1yyy}, respectively. We see
that the variational results are in perfect agreement with the numerically
exact results, confirming the validity of the variational approach.
We have verified that the variational results coincide
with the numerically exact solutions of the non-Hermitian JC model
ranging from the single to multi-mode case in a broad range of parameters. These findings suggest
that the variational approach is capable of solving TDSE governed by a non-Hermitian Hamiltonian.

\subsection{Dynamics of population decay in the non-Hermitian HTC model}\label{Om}

The HTC model is coined by augmenting the Tavis-Cummings (TC) model with a bath of phonon modes\cite{D2_TC}. The HTC model has been extensively employed in describing many-body quantum dynamics in various molecular cavity QED systems \cite{SF1,SF2}. The Hamiltonian of the HTC model can be written as
\begin{equation}\label{HTC}
\hat{H}_{\rm HTC}=\hat{H}_{\rm TC}+\hat{H}_{\rm R}+\hat{H}_{\rm I},
\end{equation}
where $\hat{H}_{\rm R}$ is the reservoir Hamiltonian, $\hat{H}_{\rm I}$ is the interaction Hamiltonian of the reservoir and the TC model, and $\hat{H}_{\rm TC}$ is
the Hamiltonian of the TC model, consisting of N two-level atoms and one cavity-mode, which reads
\begin{align}\label{TC}
\hat{H}_{\rm TC}=\omega_c\hat{a}^{\dagger}\hat{a}+\sum_{n=1}^{N}[\omega_n\hat{\sigma}_n^{+}\hat{\sigma}_n^{-}+
\frac{\omega_{\rm R}}{\sqrt{N}}(\hat{a}^{\dagger}\hat{\sigma}_n^-+\hat{a}\hat{\sigma}_n^{+})].
\end{align}
Here, $\omega_c$ is the frequency of the cavity mode, and $\omega_n$ is the transition frequency of the $n$th qubit. For simplicity, we set $\omega_n=\omega_0$. The qubit-cavity coupling is assumed to be $\omega_{\rm R}/\sqrt{N}$. {It is noted that by including the counter-rotating terms in Eq.~(\ref{TC}), TC model can be transformed into the Dicke model.}
The reservoir Hamiltonian $\hat{H}_{\rm R}$ takes the form
\begin{equation}\label{HR}
\hat{H}_{\rm R}=\sum_k\omega_k\hat{b}_k^{\dagger}\hat{b}_k,
\end{equation}
and the qubit-reservoir interaction Hamiltonian
\begin{equation}\label{HR}
\hat{H}_{\rm I}=-\frac{\lambda}{\sqrt{N}}\sum_k\sum_{n=1}^N\omega_k\hat{\sigma}_n^{+}\hat{\sigma}_n^{-}(e^{-ikn}\hat{b}_k^{\dagger}+e^{ikn}\hat{b}_k).
\end{equation}
Here, only the diagonal qubit-reservoir coupling strength $\lambda$ is considered, and a linear phonon dispersion is assumed:
\begin{equation}
\omega_k=\omega_{k0}[1+\Omega(\frac{2|k|}{\pi}-1)],
\end{equation}
where $\omega_{k0}$ denotes the central energy of the phonon band, $\Omega\in[0,1]$ is the band width, and the momentum is set to be $k=2\pi{l}/N$ with $(l=-\frac{N}{2}+1,\cdots,\frac{N}{2})$. 

The Hamiltonian in Eq.~(\ref{TC}) is Hermitian. To take into account various dissipative effects that are not included in the reservoir Hamiltonian $\hat{H}_{\rm R}$, such as cavity loss, a non-Hermitian term,
\begin{equation}
\hat{H}_{\rm nH} = -i\kappa|g, \{1\}\rangle\langle g, \{1\}|,
\end{equation}
can be added to Eq.~(\ref{TC}), where $\kappa$ is the loss rate. The effective Hamiltonian with the non-Hermitian term neglects, in comparison with the Lindblad master equation, the fluctuation term $2\kappa a\rho a^{\dagger}$.  A more sophisticated scheme includes the coupling between the cavity mode and the environmental modes (namely, the Gardiner-Collett interaction Hamiltonian), characterized by a continuum spectral density. This Hamiltonian can be rigorously derived from the QED first-principles. It has been used to investigate polariton quantum dynamics in a dissipative cavity~\cite{CR}. By constructing an effective non-Hermitian Hamiltonian or employing spectral discretization techniques, the time-dependent variation method with the mD2 Ansatz can be applied to account for dissipation scenarios.

For a nonzero cavity loss rate, both the upper and lower polariton states experience a depletion of their population, transferring it to the ground state. This loss of the excited state population typically diminishes the capacity of the system to undergo reactions on excited surfaces.

\begin{figure}[t]
\begin{minipage}[t]{0.8\linewidth}
\centering
\includegraphics[scale=0.5]{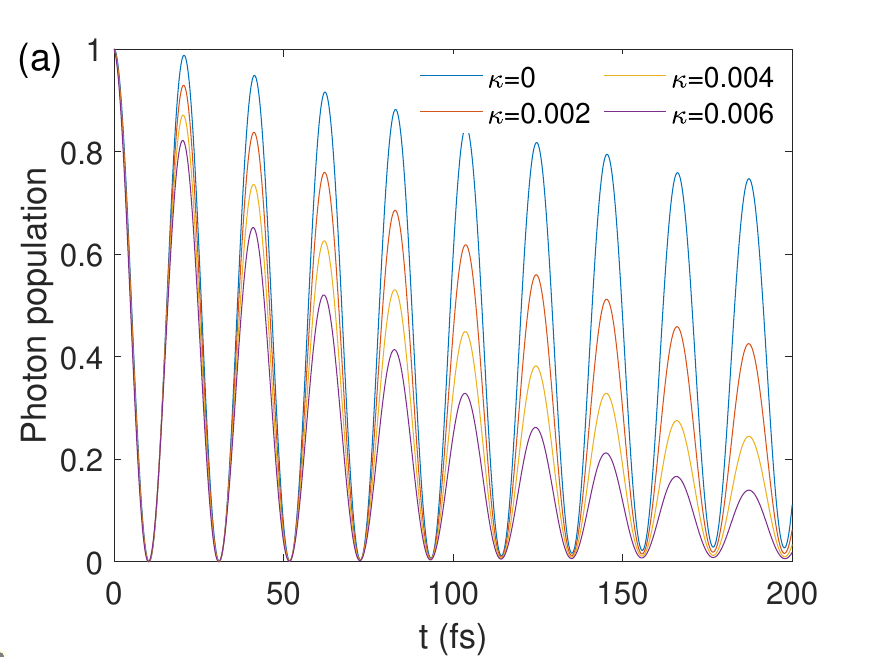}\\
\end{minipage}
\begin{minipage}[t]{0.8\linewidth}
\centering
\includegraphics[scale=0.5]{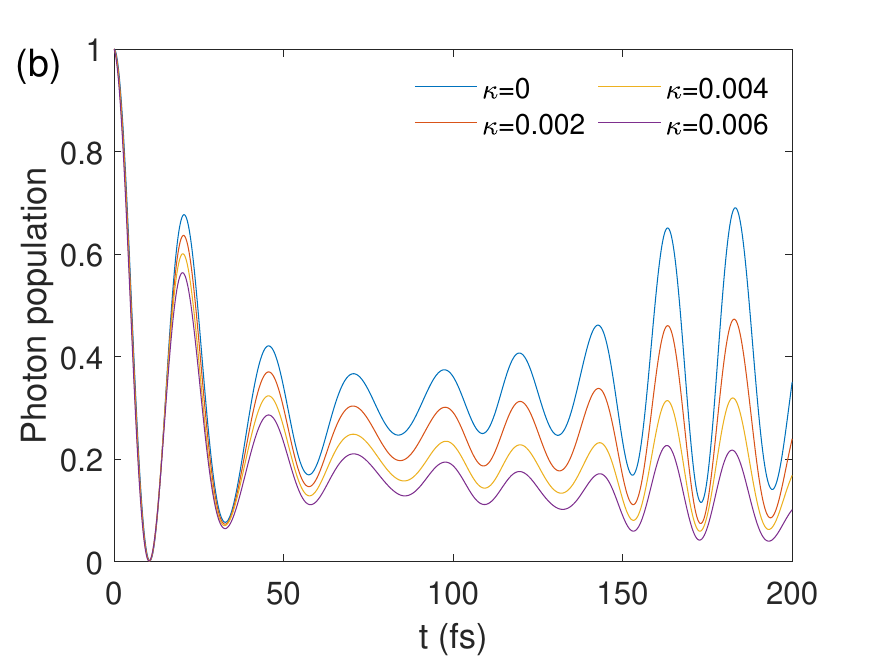}\\
\end{minipage}
\caption{The photon-mode population dynamics with the finite {inverse} lifetimes $\kappa=0\rm \;eV, 0.002\rm \;eV, 0.004\rm \;eV,$ and 0.006 eV for (a): $\lambda = 0.1$  and (b): $\lambda = 0.4$. The other parameters are set by $\omega_R=0.1\rm eV$, $N=10$, $\omega_c=\omega_n=1\rm eV$, and $\Omega=0.5$.}
\label{fig4}
\end{figure}

The photon population evolutions $\langle {{\rm D}_{2}^M(t)}|\hat{a}^{\dagger}\hat{a}| {{\rm D}_{2}^M(t)}\rangle$ for different cavity loss parameters $\kappa$ and qubit-phonon couplings $\lambda$ are plotted in Fig.~\ref{fig4}. As the cavity loss increases, the photon population gradually decreases. Due to the finite lifetimes ($\kappa\neq0$), the population decays to
zero in the long-time limit. The bright-state energy gap for TC model is $\epsilon=\sqrt{(\omega_c-\omega_0)^2+4\omega_R^2}=2\omega_R$, which determines the oscillation period $2\pi/2\omega_R\sim 20.7~\rm fs$ of the photon population. As a comparison, for the case of weaker qubit-phonon coupling in HTC model as shown in Fig.~\ref{fig4}(a), we can still observe the regular 20.7~fs periodic oscillations. For $\lambda=0.4$, the phonon-qubit coupling significantly reduces amplitudes of these oscillations, as shown in  Fig.~\ref{fig4}(b). The stronger qubit-phonon coupling quenches Rabi oscillations, due to the increasing number of quantum states participating in the HTC dynamics. It is worth mentioning that a partial recurrence of photon population occurs after $150\rm fs$. We attribute it to the limited number of discrete phonon modes in the model. In addition, the period of Rabi oscillations has a small increment for a stronger phonon-qubit coupling in Fig.~\ref{fig4}(b). Hence the renormalization Rabi frequency is slightly reduced by the qubit-phonon coupling.

\section{Conclusion}

In this work, by employing the mD2 Ansatz, we have successfully carried out numerically ``exact'' simulations of various of non-Hermitian quantum systems.  For the non-Hermitian dissipative Landau-Zener transitions, we have predicted the logarithmic {correspondence} between the unnormalized population and the strength of non-Hermicity for both single mode and multi mode scenarios, as the mD2 results coincide with that obtained via the TDSE. For the non-Hermitian, multi-mode JC model, we have found exact overlapping of the excited-state population calculated by mD2 Ansatz and the numerically exact approaches. Lastly, for the HTC model, we have demonstrate that the mD2 Ansatz can also handle Hermitian systems with non-Hermitian additions that serve as a phenomenological description of realistic experiments.

Beyond the three systems studied in this work, there are myriad other applications of mDA to non-Hermitian systems. As demonstrated in this work, mDA calculations are capable of providing phenomenalogical descriptions in non-Hermitian physics. It also serves as a powerful numerical apparatus to deal with complex many-body dynamics in non-Hermitian systems. The investigation of these systems could provide deeper insights into a range of intriguing phenomena, including non-Hermitian skin effects, non-Hermitian excited-state dynamics, and non-Hermitian evolution of quantum photonics systems. Additionally, since the mDA approach enables direct time-dependent calculations for non-Hermitian systems, it offers a convenient tool for exploring non-Hermitian spectral topology\cite{ccls2}.

\section*{Acknowledgments}
The authors thank Lu Wang, Jiarui Zeng, Qinghu Chen, and Chenlin Ma for useful discussion. M. F. G. acknowledges support from the National Natural Science Foundation of China (Grant No.~22373028). Support from the Singapore Ministry of Education Academic Research Fund Tier 1 (Grant Nos.~RG87/20 and RG2/24) is also gratefully acknowledged.

\section*{Author Declarations}

\subsection*{Conflict of Interest}
The authors have no conflicts to disclose.

\section*{Data Availability}
The data that support the findings of this study are available from the corresponding author upon reasonable request.

\appendix*
\section{Derivation of Norm Decay in Non-Hermitian Hamiltonian Systems}
We begin with a general non-Hermitian Hamiltonian
\begin{equation}\label{GH}
\hat{H} = \hat{H}_0 - i \hat{\Gamma}
\end{equation}
where $\hat{H}_0$ is Hermitian ($\hat{H}_0^{\dagger} = \hat{H}_0$), and $\hat{\Gamma}$ is a Hermitian, positive semi-definite operator ($\hat{\Gamma}^{\dagger} = \hat{\Gamma}$, $\langle \Psi(t) |\Gamma | \Psi(t) \rangle \textgreater 0$) which is responsible for the dissipative effects.

The TDSE is given by
\begin{equation}\label{Ha}
i \frac{d}{dt} | \Psi(t) \rangle = \hat{H} | \Psi(t) \rangle
\end{equation}
The norm of the wavefunction is defined as
\begin{equation}
N(t) = || \Psi(t)||^2 = \langle \Psi(t) | \Psi(t) \rangle
\end{equation}
Taking the time derivative of $N(t)$, {we obtain}
\begin{equation}\label{Dt}
\frac{d}{dt} N(t) = \frac{d}{dt} \langle \Psi(t) | \Psi(t) \rangle = \left( \frac{d}{dt} \langle \Psi(t) | \right) | \Psi(t) \rangle + \langle \Psi(t) | \frac{d}{dt} | \Psi(t) \rangle
\end{equation}
Using the adjoint form of Eq.~(\ref{Ha}), we have
\begin{equation}
\left( \frac{d}{dt} \langle \Psi(t) | \right) =  i \langle \Psi(t) | \hat{H}^\dagger
\end{equation}
Substituting this into Eq.~(\ref{Dt}), we get
\begin{align}
\frac{d}{dt} N(t) = i \left( \langle \Psi(t) | \hat{H}^{\dagger} | \Psi(t) \rangle - \langle \Psi(t) | \hat{H} | \Psi(t) \rangle \right)
\end{align}
Applying the condition
\begin{equation}
\hat{H} - \hat{H}^\dagger = (\hat{H}_0 - i \hat{\Gamma}) - (\hat{H}_0 + i \hat{\Gamma}) = -2 i \hat{\Gamma},
\end{equation}
we obtain
\begin{equation}
\frac{d}{dt} N(t) = - 2\langle \Psi(t) | \hat{\Gamma} | \Psi(t) \rangle \textless 0
\end{equation}
Thus, the total population decays with time.


\end{document}